\newif\ifdraft
\drafttrue
\draftfalse

\ifdraft
 \documentclass[draftcls,onecolumn,12pt,a4paper]{IEEEtran}
\else
	\documentclass[lettersize,journal]{IEEEtran}
\fi

\newlength\mywidth
\ifdraft
\else
\fi
\advance\mywidth by -2ex 

\usepackage{amsmath,amsfonts,amssymb}
\usepackage{algorithm}
\usepackage{array}
\usepackage[caption=false,font=normalsize,labelfont=sf,textfont=sf]{subfig}
\usepackage{textcomp}
\usepackage{stfloats}
\usepackage{url}
\usepackage{verbatim}
\usepackage{graphicx}
\usepackage{cite}
\usepackage[commentColor=mygreen]{algpseudocodex} 
\usepackage{float}
\newfloat{algorithm}{tbp}{loa} 
\floatname{algorithm}{Algorithm}

\usepackage{color} 
\definecolor{mygreen}{rgb}{0.4,0.6,0.4}

\def\Figu#1{{\bf Figure~\ref{#1}}}

\def\Tabl#1{{\bf Table~\ref{#1}}}

\def\Algo#1{{\bf Algorithm~\ref{#1}}}
\def\tdot{\!\cdot\!}

\usepackage{tikz}

\newcommand\submittedtext{%
  \footnotesize This work has been submitted to the IEEE for possible publication. Copyright may be transferred without notice, after which this version may no longer be accessible.}

\newcommand\submittednotice{%
\begin{tikzpicture}[remember picture,overlay]
	\node[anchor=north,yshift=-10pt] at (current page.north) {\fbox{\parbox{\dimexpr0.95\textwidth-\fboxsep-\fboxrule\relax}{\submittedtext}}};
\end{tikzpicture}%
	\vspace{-8mm}
}

\newcommand\copyrighttext{%
  \footnotesize \textcopyright \the\year{} IEEE. Personal use of this material is permitted. Permission from IEEE must be obtained for all other uses, including reprinting/republishing this material for advertising or promotional purposes, collecting new collected works for resale or redistribution to servers or lists, or reuse of any copyrighted component of this work in other works.}

\begin{document}

\title{Investigations on Algorithm Selection for Interval-Based Coding Methods}

\author{Tilo Strutz and Nico Schreiber
	\thanks{This work has been funded by the Deutsche Forschungsgemeinschaft (DFG, German Research Foundation) - 438221930.}
	\thanks{T. Strutz and N. Schreiber are with Coburg University, Germany.}%
}%



\maketitle
\submittednotice		

\begin{abstract}
There is a class of entropy-coding methods which do not substitute symbols by code words (such as Huffman coding), but operate on intervals or ranges. This class includes three prominent members: conventional arithmetic coding, range coding, and coding based on asymmetric numeral systems. To determine the correct symbol in the decoder, each of these methods requires the comparison of a state variable with subinterval boundaries. In adaptive operation, considering varying symbol statistics, an array of interval boundaries must additionally be kept up to date. 
The larger the symbol alphabet, the more time-consuming both the search for the correct subinterval and the updating of interval borders become. 

Detailed pseudo-code is used to discuss different approaches to speed up the symbol search in the decoder and the adaptation of the array of interval borders, both depending on the chosen alphabet size.
It is shown that reducing the $\mathcal{O}$-complexity does not lead to an acceleration in practical implementations if the alphabet size is too small. 
In adaptive compression mode, the binary indexing method proves to be superior when considering the overall processing time. Although the symbol search (in the decoder) takes longer than with other algorithms, the faster updating of the array of interval borders more than compensates for this disadvantage.
A variant of the binary indexing method is proposed, which is more flexible and has a partially lower complexity than the original approach.
\end{abstract}


\section{Introduction}
\label{sec_Introduction}
	%
\IEEEPARstart{E}{ntropy} coding is a fundamental processing step in any efficient data compression system. Based on the given or estimated probabilities $p(s_i)$ of the symbols $s_i$ to be transmitted, entropy coding assigns a certain number or fraction of bits to each symbol $s_i$. Ideally, the number of bits per symbol corresponds to its information content $I(s_i) = \log_2[1/p(s_i)]$.
 
This assignment can be achieved by the class of interval-based coding methods. One of the first well-studied methods of this class is the (conventional) {\em arithmetic coding} \cite{Ris79}. Arithmetic coding (AC) uses a number interval that is subdivided into subintervals. The relative width of these subintervals corresponds to the symbol probabilities. A second method is called {\em range coding}. The idea of range coding (RC) also dates back to the 1970s \cite{Mar79}, and on closer inspection it becomes apparent that these two methods are closely related in terms of their practical implementation in integer arithmetic. While conventional arithmetic coding operates with bits that are sent or received, range coding focuses on digits of number systems with arbitrary bases that are not necessarily equal to 2 or 10. Therefore, conventional arithmetic coding can be regarded as a special case of range coding. For convenience, a base $b=256$ is usually selected for range coding, which allows byte-wise input and output of data. The third method considered in this paper is based on asymmetric numeral systems (ANS) \cite{Dud14}. We refer here specifically to a variant known as range ANS (rANS). It also operates with bytes like RC. However, it is a first-in-last-out technique, in other words the encoder has to process the data in reverse order compared to the decoder. This complicates data handling, especially in adaptive coding settings \cite{Str23}.

All three of these interval-based coding methods can handle multi-symbol alphabets with $K\ge 2$, which has the advantage that symbols do not have to be converted into a sequence of bits requiring individual probability estimates, as is required, for example, in context-based adaptive binary arithmetic coding (CABAC) \cite{Mar03}.

The coding of large alphabets can be advantageous in many applications and is the subject of research \cite{Rya03,Che20,Str23,Och24}. 

These multi-symbol methods also have in common that decoding needs a specific search routine to find the correct symbol. Searching for alphabets with $K\gg 2$ becomes a time-consuming issue and special strategies are required to minimise time complexity.
When operating in adaptive mode, i.e. encoder and decoder learn the symbol statistics on-the-fly, an array of cumulative counts has additionally to be updated after processing each individual symbol \cite{Mof99}.

This paper provides the following contributions:
\begin{itemize}
	\item a discussion and comparison of different search algorithms in terms of their average-case and worst-case $\mathcal{O}$-complexity in application to interval-based decoding,
	\item an examination of theses search algorithms regarding their decoding time required in a real range-coder implementation, 
	\item an explanatory discussion of a working implementation of the binary indexing method (applied to interval-based coding) that is more consistent than the original text of \cite{Fen94},
	\item a new rescaling procedure based on binary indexing that is less complex than the procedure shown in \cite{Fen94}, and
	\item a proposal for selecting the best algorithms for different coding conditions.
\end{itemize}
	%
\section{Basics}\label{sec_basics}
	%
The coding procedures are based on an interval $[0;M)$, which is divided into $K$ subintervals, where $K$ is the number of different symbols $s_i$ (size of the alphabet) with $i=0,1,\dots,K-1$. 
Each subinterval corresponds to one symbol of the alphabet used.
The size of an individual subinterval is proportional to the frequency (count) of the symbols $h(s_i)$ or $h[i]$.
The cumulative counts are therefore the interval boundaries:
	\begin{align}\label{eq_sumCounts}
		h_{\rm k}[i+1] &= h_{\rm k}[i] + h[i] \quad\forall i=0,1,\dots,K-1\\
			&\quad\mbox{with}\quad h_{\rm k}[0]=0
			\quad\mbox{and}\quad  h_{\rm k}[K] = totalCount    \nonumber
	\;.
	\end{align}
That is, $h_{\rm k}[i]$ is the lower border and $h_{\rm k}[i+1]$ is the upper interval border of symbol $s_i$.

All three coding techniques mentioned have in common that a code value $c$ is derived from the stream of compressed data on the decoder side, which decides on the current subinterval and thus on the decoded symbol.
In practical terms, this code value must be compared with the interval boundaries. If $c \in [h_{\rm k}[i], h_{\rm k}[i+1])$ holds, then $s_i$ is the decoded symbol.

The cumulative count valuess $h_{\rm k}[]$ also reflect the symbol distribution. In static coding, the statistics are determined once on encoder side and then transmitted to the decoder. In adaptive mode, the symbol distribution typically starts with an assumed flat distribution and a specific count value $h[i]$ is incremented after the corresponding symbol $s_i$ has been encoded or decoded.

According to equation (\ref{eq_sumCounts}), all cumulative counts $h_{\rm k}[j]$ with $i<j \le K$ have to be increased if the count $h[i]$ of the symbol index $i$ has been incremented by one. In worst case, all $K$ cumulative counts must be considered.
For large alphabets, this updating process can be the limiting factor in terms of processing speed, so a more clever solution is needed.
In contrast to symbol search, this adaptation concerns both, encoder and decoder, as the cumulative counts have to be updated synchronously. 

It should be noted that in static compression mode, the counts $h[i]$ may be zero, because it is known in advance which symbol can occur and which cannot. In adaptive compression mode, the counts must be at least equal to one, as it must be possible to differentiate between all symbols that could occur. I other words, the cumulative counts $h_{\rm k}[i]$ may not be identical in adaptive mode; the interval width must not be equal to zero. 
	%
\section{Conventional searching methods}\label{sec_conventionalSearch}
	%
Linear search is the simplest method for the decoder to find the correct subinterval to identify the decoded symbol. It compares the code value $c$ sequentially with each interval boundary until a match is found. If the search starts at the highest index, this is referred to as a linear backward search, see \Algo{alg_getSymbolLinearBackward}.
\begin{algorithm}
 \caption{\label{alg_getSymbolLinearBackward}Determination of the correct symbol index $i$ using the linear backward search}
 \hfil
  \begin{minipage}[t]{\mywidth}
		\ifdraft
			\setlength{\baselineskip}{12pt} 
		\fi
		\begin{algorithmic}[1]
		\Procedure{getSymbolLinBackward}{$c, h_{\rm k}[~]$}
			\State $i \gets K-1$
			\Comment{start search at last symbol $s_0$}
			\While {$c < h_{\rm k}[i]$)}
				\Comment{if not inside the current sub-interval \dots}
				\State $i \gets i-1$
				\Comment{\dots go to next index}
			\EndWhile
			\State \Return $i$
		\EndProcedure	
		\end{algorithmic}
	\end{minipage}
\end{algorithm}
If symbols with high indices dominate the distribution, this search can be fast. In worst case, however, all boundaries have to be checked and the decoding process of a single symbol gets $\mathcal{O}(K)$ complexity.

In principal, the search can also start with the first symbol, \Algo{alg_getSymbolLinearForward}.
\begin{algorithm}
 \caption{\label{alg_getSymbolLinearForward}Determination of the correct symbol index $i$ using linear forward search}
 \hfil
  \begin{minipage}[t]{\mywidth}
		\ifdraft
			\setlength{\baselineskip}{12pt} 
		\fi
		\begin{algorithmic}[1]
		\Procedure{getSymbolLinForward}{$c, h_{\rm k}[~]$}
			\State $i \gets 1$
			\Comment{start search at first symbol $s_0$}
			\While {$c \ge h_{\rm k}[i]$)}
				\Comment{if not inside the current sub-interval \dots}
				\State $i \gets i+1$
				\Comment{\dots go to next index}
			\EndWhile
			\State \Return $i-1$
		\EndProcedure	
		\end{algorithmic}
	\end{minipage}
\end{algorithm}
This variant was used in the first publication of a practical source code in \cite{Wit87} and is preferable when symbols with a low index have the highest probabilities.

The worst-case complexity can be reduced to $\mathcal{O}(\log{K})$ by a logarithmic search, see \Algo{alg_getSymbolLOG}.
	    %
\begin{algorithm}
 \caption{\label{alg_getSymbolLOG}Determination of the correct symbol index $i$ using logarithmic search}
 \hfil
  \begin{minipage}[t]{\mywidth}
		\ifdraft
			\setlength{\baselineskip}{12pt} 
		\fi
		\begin{algorithmic}[1]
		\Procedure{getSymbolLogarithmic}{$c, h_{\rm k}[~]$}
			\State $bottom \gets 0; top \gets K$
			\Comment{outer interval boundaries}
			\Repeat 
				\State $i \gets (top + bottom) >> 1$
				\Comment{interval in the middle}
				\If {$(	c < h_{\rm k}[i])$}
					\State  $top \gets i$ \Comment{choose lower half}
				\Else
					\State $bottom \gets i+1$ 	\Comment{choose upper half}
				\EndIf
			\Until {($top == bottom$)}
			\Comment{stop if no further subdivision is possible}
			\State \Return $bottom - 1$
		\EndProcedure	
		\end{algorithmic}
	\end{minipage}
\end{algorithm}
In a recently published preprint, this method is called `bisection search', \cite{Sai23}. Each comparison with an interval boundary halves the number of remaining comparisons.

There are different realizations of the logarithmic search in terms of variables involved and order of computation. The variant presented in Algorithm \ref{alg_getSymbolLOG} allows the fastest executable code in our experiments.

\section{Variants of logarithmic search}
	%
	%
\subsection{Logarithmic search with optimized range split}
	%
In principal, the logarithmic search requires a constant number of steps until the correct symbol is found, regardless of the true distribution of the symbols.
The procedure assumes that each comparison splits the current interval into two new intervals that have the same probability to occur.
This is optimal if the symbols are actually evenly distributed.
In cases where the distribution is somehow skewed, it would be preferable to split the subintervals at a boundary where the sum of counts on both sides of this boundary is almost equal, that is, both sides are roughly equally likely.
Instead of calculating $i \gets \lfloor(top + bottom) /2\rfloor$ as in line 4 of Algorithm \ref{alg_getSymbolLOG}, the next interval $i$ should be chosen such that 
 $h_{\rm k}[i]\approx (h_{\rm k}[bottom] + h_{\rm k}[top]) / 2$ holds, i.e. it must be determined according to
	\begin{align}\label{eq_bestIntervalSplit}
		i = \mathop{\arg \min}\limits_{j}\left(\left| 2\cdot h_{\rm k}[j]
					 - h_{\rm k}[top] - h_{\rm k}[bottom] \right| \vphantom{\binom{1}{1}}\right)
					\;.
	\end{align}

\begin{figure*}
	\hfil \includegraphics[scale = 0.3]{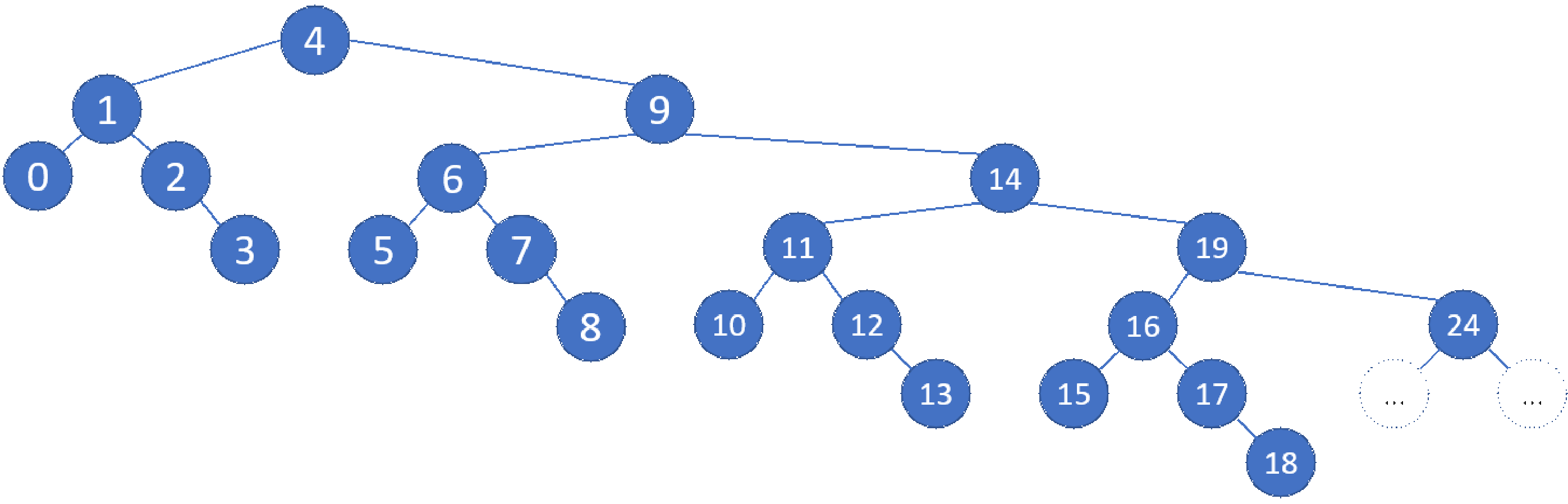}
	\caption{\label{fig_binarySearchTree}Example of binary search tree (cut-out) reflecting a geometrical distribution (skewed towards small indices)}
\end{figure*}

Based on this decision, the order of the intervals to be tested can be predetermined at the encoder and represented as a binary search tree.
\Figu{fig_binarySearchTree} shows an example tree for geometrically distributed data. 
According to the skewed distribution, the tree is unbalanced. The top index has been set to 4, since the indices $0\dots 3$ have approximately the same probability as the indices $5\dots K-1$ in this example.
The search tree for purely logarithmic search would be a balanced tree.
\Algo{alg_logarithmicPath} shows the decoding routine using such a search tree.
\begin{algorithm}
 \caption{\label{alg_logarithmicPath}Logarithmic search with optimized range split stored in the arrays $le\!f\!t[]$ and $right[]$}
 \hfil
  \begin{minipage}[t]{\mywidth}
		\ifdraft
			\setlength{\baselineskip}{12pt} 
		\fi
		\begin{algorithmic}[1]
		\Procedure{getSymbolTree}{ $c, h_{\rm k}[]$, $i_{\rm Mid}$, $le\!f\!t[]$, $right[]$}
			\State $i \gets i_{\rm Mid}$
			\Comment{the top index is separately given}
			\While{true} 
				\If {$(	c < h_{\rm k}[i])$}
					\State  $i \gets le\!f\!t[i]$ \Comment{goto left child}
				\Else
					\If {$(	c < h_{\rm k}[i+1])$}
						\State{\bf break} \Comment {index found, leave loop}
					\EndIf
					\State $i \gets right[i]$ \Comment{goto right child}
				\EndIf
			\EndWhile
			\State \Return $i$
		\EndProcedure	
		\end{algorithmic}
	\end{minipage}
\end{algorithm}
The arrays $le\!f\!t[]$ and $right[]$ must be prepared on the basis of (\ref{eq_bestIntervalSplit}) and contain the child indices as illustrated in Figure \ref{fig_binarySearchTree}. 

In adaptive mode, one could start with a balanced tree. Based on the collected statistics (i.e. the updated symbol counts) the search tree needs to be reorganized by rotation.
Unfortunately, this reorganization usually takes more time than is saved by reducing the number of iterations.

Therefore, for the adaptive mode, we suggest to determine the best index only for the initial range split and then continue with the standard logarithmic search. This initial value is either determined once in static mode using procedure {\scshape  determineInitialSplit()} or it starts with $i_{\rm Mid} \gets \lfloor K/2\rfloor$ and can be easily adapted based on the true symbol index as shown in {\scshape  adaptInitialSplit()} of \Algo{alg_logarithmic2}.
\begin{algorithm}
 \caption{\label{alg_logarithmic2}Logarithmic search with adapted initial index}
 \hfil
  \begin{minipage}[t]{\mywidth}
		\ifdraft
			\setlength{\baselineskip}{12pt} 
		\fi
		\begin{algorithmic}[1]
		\Procedure{getSymbolLog2}{$K, c, h_{\rm k}[~], i_{\rm Mid}$}
			\State $bottom \gets 0; top \gets K$
			\Comment{outer interval boundaries}
			\State $i \gets i_{\rm Mid}$
			\Repeat 
				\If {$(	c < h_{\rm k}[i])$}
					\State  $top \gets i$ \Comment{choose lower half}
				\Else
					\State $bottom \gets i+1$ 	\Comment{choose upper half}
				\EndIf
				\State $i \gets (top + bottom) >> 1$
				\Comment{interval in the middle}
			\Until {($top == bottom$)}
			\Comment{stop if no further subdivision is possible}
			\State \Return $bottom - 1$
		\EndProcedure	
		\State {}
		\Procedure{determineInitialSplit}{$K, h_{\rm k}[~]$}
			\State $i_{\rm Mid} = 0$
			\Comment{find index that halves sum of all counts}
			\While {$h_{\rm k}[i_{\rm Mid}] < h_{\rm k}[K]/2$}
					\State $i_{\rm Mid} \gets i_{\rm Mid} + 1$
			\EndWhile
			\If {$(h_{\rm k}[i_{\rm Mid}] > ( h_{\rm k}[K] - h_{\rm k}[i_{\rm Mid}-1]))$}
					\State $i_{\rm Mid} \gets i_{\rm Mid} -1$
					\Comment{choose better candidate of two neighbours}
			\EndIf
			\State \Return $i_{\rm Mid}$
		\EndProcedure	
		\State {}
		\Procedure{adaptInitialSplit}{$K, i_{\rm Mid}, i$}
			\If {$i_{\rm Mid} < i$}
			\Comment{adapt initial split based on true index}
				\If {$i_{\rm Mid} > 0$}
					\State $i_{\rm Mid} \gets i_{\rm Mid} - 1$
				\EndIf
			\Else
				\If {$i_{\rm Mid} < K$}
					\State $i_{\rm Mid} \gets i_{\rm Mid} +1$
				\EndIf
			\EndIf
			\State \Return $i_{\rm Mid}$
		\EndProcedure	
		\end{algorithmic}
	\end{minipage}
\end{algorithm}

The effectiveness of such a scheme can be verified by counting the number of iterations until the target index is found.
For an alphabet with $K=64$ symbols following a truncated geometric distribution, we measured the probabilities for the number of iterations in percent, \Tabl{tab_numberOfIterations}.
\begin{table*}
	\caption{\label{tab_numberOfIterations}Number of iterations (in percentage) needed for the identification of one out of $K\!=\!64$ geometrically distributed symbols, depending on the chosen algorithm}
	\hfil
	\begin{tabular}{|c|*{15}{c}|c|}
	\hline
	\# iter.:	&	1			&	2			&	3			&	4		&	5			&	6			&	7			&	8			&	9			&	10	&	11		&	12		&	13		&	14	&	15	&ave.		\\
	\hline
	LOG 	& 0.00 	& 0.00 	& 0.00 	& 0.00 & 0.00 & 84.09 & 15.91 & 0.00 	& 0.00 	& 0.00 & 0.00 & 0.00 & 0.00 & 0.00	& 0.00	& 6.2  \\
	TREE 	& 7.96	&16.72	&34.20	&23.84	&10.02&	4.22	&	1.77	&	0.75	&0.31		&	0.13 & 0.06	&	0.02 & 0.01	&4.5e-03	&8.3e-04	& 3.4\\
	LOG2	& 0.00 	& 0.00 	& 34.10	&15.91	&0.00	&	0.76	&49.23	&	0.00 	& 0.00 	& 0.00 & 0.00 & 0.00 & 0.00 & 0.00 	&	0.00	&  5.2 \\
	\hline
	\end{tabular}
\end{table*}
In a purely logarithmic search (row LOG), one would expect $\log_2(64)=6$ iterations for each symbol. However, there are around 16 percent of symbols that need seven iterations. This additional iteration is required to identify the symbol index zero.
When using the optimized binary search tree (row TREE), about eight percent of the symbols need only a single iteration. This is the case, when the top index is the correct one. More than 50 percent of the symbols require only three or four iteration, reducing the average number of iterations from 6.2 to 3.4. The compromise according to Algorithm \ref{alg_logarithmic2} (row LOG2) has an average of 5.2 iterations.
	%
\subsection{Exponential search}
	%
A computationally more efficient method to improve the logarithmic search is to combine it with a preselection of the interval range, as shown in \Algo{alg_exponentialSearch}, lines 2-5.
\begin{algorithm}
 \caption{\label{alg_exponentialSearch}Logarithmic search in combination with pre-selection of the desired range}
 \hfil
  \begin{minipage}[t]{\mywidth}
		\ifdraft
			\setlength{\baselineskip}{12pt} 
		\fi
		\begin{algorithmic}[1]
		\Procedure{getSymbolExponential}{$c, h_{\rm k}[~]$}
			\State $bottom \gets 0; top \gets 1$
			\Comment{initial interval borders}
			\While {$h_{\rm k}[top] <= c$} \Comment{look for needed range}
				\State $top = top << 1$	\Comment {double needed range}
			\EndWhile
			\State $bottom = top >> 1$ \Comment{set lower border}
			\Repeat  \Comment {do standard logarithmic search}
				\State $i \gets (top + bottom) >> 1$
				\Comment{index in the mid}
				\If {$(	c < h_{\rm k}[i])$}
					\State  $top \gets i$ \Comment{choose lower half}
				\Else
					\State $bottom \gets i+1$ 	\Comment{choose upper half}
				\EndIf
			\Until {($top == bottom$)}
			\Comment{stop if no further subdivision is possible}
			\State \Return $bottom - 1$
		\EndProcedure	
		\end{algorithmic}
	\end{minipage}
\end{algorithm}
This so-called exponential search starts with a range $[bottom,top]=[0,1]$ and doubles this range until the code value $c$ is contained. The corresponding complexity is $\mathcal{O}(\log i)$ and in the worst case $\mathcal{O}(\log K)$. The range for the logarithmic search can now be limited to $[top/2,top]$ and has the same complexity.

If the most probable symbols have low indices $i$, then the while loop in lines 3-4 of  Algorithm \ref{alg_exponentialSearch} requires only a few iterations on average and the resulting logarithmic search range is also rather small. Overall, the complexity is $\mathcal{O}(\log i + \log i)$.
However, if symbols with large indices dominate, the while loop mentioned comes on top without accelerating anything and thus slows down the entire processing ($\mathcal{O}(\log K + \log K)$).
	%
\section{Table-based searching}\label{sec_tableBased}
	%
The iterative search for an individual subinterval can be completely avoided by using a prepared array of values that contains the corresponding symbol index for each code value $c$. The size of this array must be at least as large as the total count $h_{\rm k}[K]$ of symbols. For example, if there are four different symbols (i.e. $K=4$) with counts according to following table
\begin{center}
\begin{tabular}{c|*4{c}}
  $i$		 	& 0 &	1	&	2	&	3	\\
	\hline
	$h[i]$	&	3	&	2	&	1	&	4
\end{tabular}\;,
\end{center}
the total count is equal to 10. The code value $c$ is an element of the half-open interval $[0,h_{\rm k}[K])$ and the lookup table $t[]$ would be
\begin{center}
\begin{tabular}{c|*{10}{c}}
  $c$							& 0 &	1	&	2	&	3	&	4	&	5	&	6	&	7	&	8	&	9	\\
	\hline
	$i=t[c]$							&	0	&	0	&	0	&	1	&	1	&	2	&	3	&	3	&	3	&	3	
\end{tabular}
\;.
\end{center}
Each symbol index $i$ appears exactly $h[i]$ times in the table. The great advantage of this method is that each code value $c$ can be directly mapped to the correct symbol index $i$. 

This table-based search is the best option in static compression mode, where the symbol distribution is determined once in advance (\Algo{alg_tableSearch}, procedure {\scshape  createTable()}) and is not be altered later. 
The search complexity is $\mathcal{O}(1)$, see Algorithm \ref{alg_tableSearch}, procedure {\scshape  getSymbolTableBased()}.
\begin{algorithm}
 \caption{\label{alg_tableSearch}Required procedures of table-based search: initialization, mapping, updating}
 \hfil
  \begin{minipage}[t]{\mywidth}
		\ifdraft
			\setlength{\baselineskip}{12pt} 
		\fi
		\begin{algorithmic}[1]
		\Procedure{createTable}{$K, h[~], t[~]$}
			\State $idx \gets 0$, $i \gets 0$
			\Repeat 		\Comment {for all symbol indices}
				\State $cnt \gets 0$
				\Repeat  \Comment {write symbol index into table}
					\State $t[idx] \gets i$, $idx \gets idx +1$
					\State $cnt \gets cnt + 1$
				\Until {($cnt == h[i]$)}\Comment {repeat $h[i]$ times}
				\State $i \gets i + 1$  \Comment {next symbol index}
			\Until {($i == K$)}
			\State \Return $t[~]$
		\EndProcedure	
		\State {}
		\Procedure{getSymbolTableBased}{$c, t[~]$}
			\State $i \gets t[c]$
			\State \Return $i$
		\EndProcedure	
		\State {}
		\Procedure{updateTable}{$K, t[~], h_{\rm k}[~], i$}
			\Repeat
				\State $idx = h_{\rm k}[i+1]$ \Comment {position to be changed}
				\State $t[idx] \gets i$	\Comment {set new symbol index $i$}
				\State $i \gets i + 1$  \Comment {go to next symbol index}
			\Until {(i == K)}
			\State \Return $t[~]$
		\EndProcedure	
		\end{algorithmic}
	\end{minipage}
\end{algorithm}

In adaptive coding mode, the mapping table $t[~]$ has to be modified after each individual symbol has been processed. Since this table has the size of $h_{\rm k}[K]$, this modification seems to be very time-consuming. 
However, if one examines how the mapping table changes when a single symbol $s_i$ with a certain index $i$ has been processed, it becomes obvious that the modification complexity in the worst case is only $\mathcal{O}(K)$ instead of $\mathcal{O}(h_{\rm k}[K])$. Looking at the toy example above, let us assume that the count of $s_1$ has been incremented from 2 to $h[1] = 3$. The mapping table must then be changed to 
\begin{center}
\begin{tabular}{c|*{11}{c}}
  $c$							& 0 &	1	&	2	&	3	&	4	&	5	&	6	&	7	&	8	&	9	&10\\
	\hline
	$i=t[c]$							&	0	&	0	&	0	&	1	&	1	&	1 & 2	&	3	&	3	&	3	&	3	
\end{tabular}
\end{center}
	%
It can be seen that only $K-i=4-1=3$ table entries have been changed, namely at $c=5,6$, and $10$. This means that in the worst case only $K$ entries need to be accessed and the update complexity is indeed $\mathcal{O}(K)$. Procedure {\scshape updateTable()} in Algorithm \ref{alg_tableSearch} shows the corresponding pseudo-code. The attentive reader will probably have noticed that the mapping table now contains one more element. This means that sufficient space should be allocated for this table in advance.

Although this update process is quite efficient, its worst-case complexity $\mathcal{O}(K)$ is added to the search complexity of $\mathcal{O}(1)$ and overall it takes more time than a logarithmic search, which has a complexity of only $\mathcal{O}(\log K)$ . 
	%
\section{Updating of cumulative counts}
	%
While the process of symbol identification concerns only the decoder, the second time-consuming process, the adaptation of cumulative counts, is a matter of both the encoder and the decoder as already mentioned in Section \ref{sec_basics}.

In the following, the standard algorithm is first shown and its disadvantages are discussed. After that, a smart algorithm called binary indexing, first published in \cite{Fen94}, is explained. We present  a slight variation of the original idea and propose a new and less complex algorithm for rescaling of cumulative counts.
	%
\subsection{Linear update of cumulative counts}
	%
If the cumulative counts are stored sequentially in an array, the update can be performed with a single loop, as shown in \Algo{alg_updateSumCounts}.
\begin{algorithm}
 \caption{\label{alg_updateSumCounts}Simple update of cumulative counts (with $\mathcal{O}(K)$ complexity) after processing of ymbol $s_i$}
 \hfil
  \begin{minipage}[t]{\mywidth}
		\ifdraft
			\setlength{\baselineskip}{12pt} 
		\fi
		\begin{algorithmic}[1]
		\Procedure{updateSumCounts}{$h_{\rm k}[~], i$}
			\Repeat
				\State $i \gets i+1$
				\Comment{lower boundary of symbol must kept unchanged}
				\State $h_{\rm k}[i] \gets h_{\rm k}[i] + 1$
				
			\Until {($i == K$)}\Comment{including last entry}
			\State \Return $h_{\rm k}[~]$
		\EndProcedure	
		\end{algorithmic}
	\end{minipage}
\end{algorithm}
This algorithm is quite simple and requires only a few basic operations. When symbols $s_i$ with high index $i$ have high probabilities, only a few iterations are necessary and this procedure can be fast. However, if small indices occur more frequently, we observe the opposite. The worst-case complexity is $\mathcal{O}(K)$ and the average complexity (for symmetric distributions) is $\mathcal{O}(K/2)$.

For large alphabets, this updating process can be the limiting factor in terms of processing speed and a more clever solution is required.
	%
\subsection{Binary Indexing (BI) for fast updating}
	%
\subsubsection{General idea}
	%
The symbol search described in Section \ref{sec_conventionalSearch} could in principle be accelerated by gradually splitting the array into two parts, thus reducing the complexity from $\mathcal{O}(K)$ to $\mathcal{O}(\log K)$.
Regarding updating the cumulative counts, Peter Fenwick presented a related idea in \cite{Fen94} based on a hierarchical representation of the cumulative counts. Few other researches later adopted this concept for various compression schemes (\cite{Mof99,Rya03,Pai16,Sev22}) and other query-related problems.

We review the basic ideas of this hierarchical representation and propose a variant that can handle arbitrary alphabet sizes $K$, while the original approach in \cite{Fen94} assumed that $K$ is a power of two. In addition, we will propose a new rescaling procedure that is less complex than the algorithm of \cite{Fen94}.

In its upper part, \Tabl{tab_toyexample} shows a toy example of an alphabet with $K=19$ symbols and their corresponding counts $h[]$ and cumulative counts $h_{\rm k}[]$.
\begin{table*}
	\caption{\label{tab_toyexample}Example for the hierarchical representation of cumulative counts}
	\hfil 
	\begin{tabular}{c|*{20}c}
		        $i$    & 0		 & 1 & 2 		 & 3 & 4 		 & 5 & 6 & 7	 & 8 			& 9 & 10& 11 	& 12 	& 13 	& 14 	& 15 	& 16 		& 17 	& 18 	& $K=19$\\
		\hline                                                                                                        
		        $h[i]$ & 3 		 & 2 & 2 		 & 1 & 4 		 & 1 & 5 & 2	 & 3 			& 1 & 2 & 3 	& 1 	& 4 	& 2 	& 1 	& 1 		& 3		& 2		&	-		\\
		\hline                                                                                                        
		$h_{\rm k}[i]$ &{\bf 0}& 3 &{\bf 5}& 7 &{\bf 8}& 12& 13& 18	 &{\bf 20}& 23& 24& 26 	& 29 	& 30 	& 34 	& 36 	&{\bf 37}& 38 & 41 	&	43	\\
		\hline                                                                                                        
		level 1        &{\bf 0}&   &   		 &   &   		 &   &   &  	 &				&   &   &   	&   	&   	&   	&   	&{\bf 37}&		&			&		\\
		level 2        &$\mid$ &   &   		 &   &   		 &   &   &  	 &{\bf 20}&   &   &   	&   	&   	&   	&   	&$\mid$ &			&			&		\\
		level 3        &$\mid$ &   &   		 &   &{\bf 8}&   &   &  	 &$\mid$ 	&   &   &   	&  9 	&   	&   	&   	&$\mid$	&			&			&		\\
		level 4        &$\mid$ &   &{\bf 5}&   &$\mid$ &   & 5 &   	 &$\mid$	&   & 4 &   	&$\mid$&   	&  5	&   	&$\mid$	&			&		4	& 		\\
		level 5        &$\mid$ & 3 &$\mid$ & 2 &$\mid$ & 4 &$\mid$&5 &$\mid$	& 3 &$\mid$& 2&$\mid$&	1	&$\mid$&	2	&$\mid$	&		1 &$\mid$&	2	\\
		\hline
		$v[]$   			 & 0		 & 3 & 5 		 & 2 & 8 			& 4 & 5 & 5	& 20& 3 & 4 & 2 & 9 & 1 & 5 & 2 & 37 & 1 &	4  & 2\\
	\end{tabular}
\end{table*}
We would like to emphasize here that the cumulative counts act as borders of the subintervals and $h_{\rm k}[i]$ is the lower boundary of the subinterval of the symbol $s_i$. Therefore, $h_{\rm k}[0]$ is always equal to zero and must never be changed. Since in adaptive mode each subinterval must have a minimum size of one, i.e. $h_{\rm k}[i+1] - h_{\rm k}[i] > 0$, the symbol counts $h[i]$ must also be at least equal to one as a prerequisite. 

All cumulative counts are now expressed by a sum of values from different levels of a hierarchy. For example, we see that all $h_{\rm k}[i]$ for $i\ge 16$ are equal to or greater than 37. This means that we can subtract 37 from these cumulative counts or, in other words, $v[16]=37$ is the first summand of all $h_{\rm k}[i], i\ge 16$.
The top level position $i=16$ divides the entire range in a left part where the cumulative counts are less than 37 and a right part with $h_{\rm k}[i]\ge 37$.
The second-level position $i=8$ further divides the left part into a left sub-part where the cumulative counts are smaller than 20 and a right sub-part with $20\le h_{\rm k}[i]< 37$ and so on.

Each cumulative count is now a sum of values at the current and higher levels (towards smaller indices) of the hierarchy. 
For example, according to Table \ref{tab_toyexample} we have :
	\begin{align}
		h_{\rm k}[16] &= v[16],  \nonumber\\
		h_{\rm k}[17] &= v[16] + v[17],  \nonumber\\
		h_{\rm k}[18] &= v[16] + v[18],  \nonumber\\
		h_{\rm k}[19] &= v[16] + v[18] + v[19],  \nonumber\\
		h_{\rm k}[3] &= v[0] + v[2] + v[3], \\
		h_{\rm k}[7] &= v[0] + v[4] + v[6] + v[7],  \nonumber \\
		h_{\rm k}[8] &= v[0] + v[8],	\quad \mbox{or}	\nonumber \\
		h_{\rm k}[10] &=v[0] + v[8]  + v[10]\nonumber
		\;.
	\end{align}
The last summand is always the individual part, while the other summands are shared with other cumulative counts.

How can this hierarchy help? Let us assume that symbol $s_{15}$ has been processed. Its count $h[15]$ needs to be incremented and consequently all $h_{\rm k}[i]$ with $15<i\le K$. This can now be achieved by simply increasing the value $v[16]$ from 37 to 38. Automatically all cumulative counts to the right of it are also increased!
For example, if symbol $s_6$ has been processed, only the values $v[7]$, $v[8]$, and $v[16]$ need to be changed. The worst case concerns symbol $s_0$: one has to increment $v[1]$, $v[2]$, $v[4]$, $v[8]$, and $v[16]$.

One question that remains to be answered is why the top level does not start in the middle of the entire range (i.e. at $i = \lfloor K/2\rfloor= \lfloor 19/2\rfloor =9$) but at $i=16$. Fenwick found a clever way to address the required values $v[]$ based on a binary scheme. Therefore, we first need to find the smallest power of 2 that is greater than $K$, and then take half of that for the top-level index:
	\begin{align}
		topLevIndex = \min_k \left(2^{k-1}\right) \;|\; 2^k > K, k\in \mathbb{N}_+
	\end{align}
	%
	%
\subsubsection{Procedures for binary indexing}
	%
When operating with the hierarchy array $v[]$, various functions are required to either extracting the desired information or to modify this array. \Algo{alg_initSumCounts} shows three procedures which are directly related to $h_{\rm k}[i]$. 
\begin{algorithm}
 \caption{\label{alg_initSumCounts}Procedures of binary indexing: initialization of cumulative counts assuming flat distribution ($h[i]=1$ $\; \forall\; i=0,1,\dots, K-1$); computation of the cumulative count $h_{\rm k}[i]$; updating of all cumulative counts $h_{\rm k}[]$ if count of $h[s]$ has been incremented}
 \hfil
  \begin{minipage}[t]{\mywidth}
		\ifdraft
			\setlength{\baselineskip}{12pt} 
		\fi
		\begin{algorithmic}[1]
		\Procedure{initSumCountsBI}{$v[~], K$}
		\State {}
		\Comment {assume that array $v[~]$ is set to zero $\forall i$}
			\For { $sym \gets 0$ to $K-1$}\Comment{for all symbol indices}
				\State $i \gets sym+1$
				\Comment{$i=0$ is not allowed}
				\Repeat 
					\Comment{lower boundary of symbol must kept unchanged}
					\State $v[i] \gets v[i] + 1$
					\State $i \gets i \;+ \;$ BitAND $(i, -i)$
					\Comment {get next index}
				\Until {($i > K$)}\Comment{including last entry $i==K$}
			\EndFor 
			\State {$totalCount = K$}
			\Comment {each symbol has a count equal to one}
			\State
			\Comment {maintaining totalCount avoids access of $h_{\rm k}[K]$}
			\State \Return $v[~], totalCount$
		\EndProcedure	
		\State {}
		\Procedure{getSumCountBI}{$v[~], i$}
		\State {$h_{\rm k} \gets 0$}
		\While {$i>0$} 
			\Comment{if i==0, then lower border of symbol==0 is zero}
			\State $h_{\rm k} \gets h_{\rm k} + v[i]$
			\Comment {sum up all required entries of $v[~]$}
			\State {}
			\Comment {starts always with most-right value}
			\State $i \gets  \;$ BitAND($i, i-1$)
			\Comment {get next index}
		\EndWhile {}
		\State \Return {$h_{\rm k}$}
		\EndProcedure	
		\State {}
		\Procedure{updateSumCountsBI}{$v[~], s, totCnt$}
		\If {($totCnt \ge$ MAX\_TOTALCOUNT)}
			\State{}
			\Comment{check whether cumulative counts can be further increased}
			 \State {{\scshape reScaleSumCounts($v[~],K$)}}
			\Comment{downscale values if needed}
			\State
		\EndIf
		\State {$i \gets s + 1$}			\Comment {symbol index plus one}
		\Repeat 
			\State $v[i] \gets v[i] + 1$
			\Comment {increment}
			\State $i \gets i \;+ \;$ BitAND($i, -i$)
			\Comment {determine next position to be modified}
		\Until {($i > K$)}
		\State $totCnt \gets totCnt +1$
		\Comment {update total count}
		\State \Return $v[~], totCnt$
		\EndProcedure	
		\end{algorithmic}
	\end{minipage}
\end{algorithm}
With {\scshape initSumCounts()} all $h_{\rm k}[i]$ (i.e. all $v[~]$) are initialized. This function has to be called only once in the beginning. 
The procedure {\scshape getSumCount()} determines a specific $h_{\rm k}[i]$ (lower border of the subinterval) and {\scshape updateSumCounts()} modifies the array $v[]$ after a certain symbol $sym$ has been processed. 

In addition to array $v[]$, a special variable $totalCount$ is maintained. It is equal to $h_{\rm k}[K]$ and replaces frequent calls of this particular cumulative count.

In all three procedures, we see that the positions of $v[]$ are accessed via the variable $i$, which itself is modified with a BitAND operation. This operator combines the two operands bit-by-bit.
\Tabl{tab_BitAND} explains the BitAND operation based on a simple example.
\begin{table*}
\caption{\label{tab_BitAND}Example of bit-wise AND-operation}
\hfil
\begin{tabular}{|*{5}{c|}|*{3}{c|}}
	\hline
	$i$	& $(i)_2$ &$(-i)_2$ 	& $(j)_2=$ BitAND($ i, -i$) & $j$ 	& $(i-1)_2$ 	& $(k)_2=$ BitAND($ i, i-1$) & $k$ \\
	\hline
	1		&	00001		& 11111		& 00001	& 1 &	00000	&	00000	& 0\\
	2		&	00010		& 11110		& 00010	& 2 &	00001	&	00000	& 0 \\
	3		&	00011		& 11101		& 00001	& 1 &	00010	&	00001	& 2 \\
	4		&	00100		& 11100		& 00100	& 4 &	00011	&	00000	& 0 \\
	5		&	00101		& 11011		& 00001	& 1 &	00100	&	00100	& 4 \\
	6		&	00110		& 11010		& 00010	& 2 &	00101	&	00100	& 4 \\
	7		&	00111		& 11001		& 00001	& 1 &	00110	&	00110	& 6 \\
	8		&	01000		& 11000		& 01000	& 8 &	00111	&	00000	& 0 \\
	9		&	01001		& 10111		& 00001	& 1 &	01000	&	01000	& 8 \\
	10	&	01010		& 10110		& 00010	& 2 &	01001	&	01000	& 8 \\
	11	&	01011		& 10101		& 00001	& 1 &	01010	&	01010	& 10 \\
	\hline
\end{tabular}
\end{table*}

The expression $(-i)_2$ is the two's complement of the binary version of $i$. If both operands are bitwise AND-combined, the result is equal to $2^p$, when $i$ contains $p$ prime factors `2'. The result $j$ gives the distance from current position to the position of the next higher level. The bitwise combination of $i$ and $i-1$ gives directly the position of the next higher level to the left.
Thus, with this BitAnd operation, all positions of $v[~]$ that contribute to a certain value $h_{\rm k}[]$ can be quickly determined. 

The coding process also needs access to the count $h[i]$ of the symbol $s_i$. Instead of providing a dedicated array of counts, these values can also be derived from array $v[]$ as shown in \Algo{alg_getCount}.
\begin{algorithm}
 \caption{\label{alg_getCount}Derivation of a single count $h[i]$}
 \hfil
  \begin{minipage}[t]{\mywidth}
		\ifdraft
			\setlength{\baselineskip}{12pt} 
		\fi
		\begin{algorithmic}[1]
		\Procedure{getCountBI}{$v[~], sym$}
		\State {$i \gets sym + 1$} \Comment{sym is the index of symbol $s_i$}
		\State {$h \gets v[i]$}
		\Comment{initialize the count variable}
		\State {$parent \gets \;$ BitAND $(i, i-1)$}
		\Comment{get parent index}
		\State $i \gets i - 1$
		\Comment{position of predecessor}
		\While {$parent$ NotEqual $i$} 
			\State $h \gets h - v[i]$
			\Comment {subtract required entries of $v[]$}
			\State $i \gets \;$ BitAND $(i, i-1)$
			\Comment {determine next predecessor index}
		\EndWhile {}
		\State \Return $h$
		\EndProcedure	
		\end{algorithmic}
	\end{minipage}
\end{algorithm}
Note that the BitAND operator here combines $i$ and $i-1$, which is different from the other procedures.

In Algorithm \ref{alg_initSumCounts}, procedure {\scshape updateSumCountsBI()}, we have already seen that the cumulative count can only be incremented as long as the total count has not yet reached a maximum value \texttt{MAX\_TOTALCOUNT}. If the total count is too large, then all counts and thus also the cumulative counts must be downscaled. Before discussing this in Subsection \ref{subsubsec_rescaling}, we need to think a little about the function {\scshape getSymbolLogarithmic()} shown in Algorithm \ref{alg_getSymbolLOG}.

This function requires repeated comparisons with different $h_{\rm k}[i]$. The access could be achieved using {\scshape getSumCountBI()}, Algorithm \ref{alg_initSumCounts}. The total complexity would then be about $\mathcal{O}(\log K \times \log K)$.
Here too, the array $v[]$ can be used to reduce the complexity. \Algo{alg_getSymbol} shows a variant of Fenwick's original procedure to determine the symbol index based on a code value $c$.
\begin{algorithm}
 \caption{\label{alg_getSymbol}Decoder procedure for determination of the symbol index based on a code value $c$ and using the binary indexing}
 \hfil
  \begin{minipage}[t]{\mywidth}
		\ifdraft
			\setlength{\baselineskip}{12pt} 
		\fi
		\begin{algorithmic}[1]
		\Procedure{getSymbolBI}{$v[],K, topLevIdx, c$}
			\State {$bottom \gets 0$}
			\Comment {initial separation index}
			\Repeat 
				\State $testIdx \gets bottom + topLevIdx$
				\Comment{ probe midpoint of range}
				\If {($c \ge v[testIdx]$ AND $testIdx\le K$ )}
					\State $bottom \gets testIdx$
					\Comment {copy bottom to probe index}
					\State $c \gets c - v[bottom]$
					\Comment {subtract higher level value}
				\EndIf
				\State {$topLevIdx \gets topLevIdx >> 1$}
			\Until {$(topLevIdx == 0)$}\Comment{repeat until range disappears}
		\State \Return $bottom$\Comment{index of symbol found}
		\EndProcedure	
		\end{algorithmic}
	\end{minipage}
\end{algorithm}
Its complexity is solely $\mathcal{O}(\log K)$.
	%
\subsubsection{Rescaling of cumulative counts}\label{subsubsec_rescaling}
	%
The value of \texttt{MAX\_TOTALCOUNT} must be large enough to represent the symbol statistics with some precision. 
In our experiments, we chose \texttt{MAX\_TOTALCOUNT = $2^{20}$}. 
When the total count reaches this maximum, all counts must be lowered.
This rescaling is usually performed by dividing all counts $h[i]$ by two. Fenwicks rescaling procedure reflects this idea, see \Algo{alg_rescale}.
\begin{algorithm}
 \caption{\label{alg_rescale}Downscaling of cumulative counts $h_{\rm k}[i]$}
 \hfil
  \begin{minipage}[t]{\mywidth}
		\ifdraft
			\setlength{\baselineskip}{12pt} 
		\fi
		\begin{algorithmic}[1]
		\Procedure{reScaleSumCountsBI}{$v[~], K$}
			\For { $sym \gets 0$ to $K-1$}\Comment{for all symbols}
				\State {$h \gets \;$ {\scshape getCount}($v[], sym$)}
				\Comment {get count that has to be halved}
				\State {$h \gets h >> 1$}
				\State $i \gets sym+1$
				\Comment {set first position in $v[]$}
				\Repeat 
					\State $v[i] \gets v[i] - h$
					\Comment{subtract half count value}
					\State $i \gets i \;+ \;$ BitAND $(i,-i)$
					\Comment {determine next position}
				\Until {($i > K$)}\Comment{including last entry $i==K$}
			\EndFor 
			\State $totalCount = \;${\scshape getSumCount}($v[], K$)
		\State \Return $v[], totalCount$
		\EndProcedure	
		\end{algorithmic}
	\end{minipage}
\end{algorithm}
For each symbol, the corresponding count is first determined and halved. Then all associated entries of $v[]$ are modified. It could be observed that the function {\scshape getCount()} requires $2K$ accesses to $v[]$ for each individual symbol. Changing the array $v[]$ requires a further $2K+[\log_2(K) -2]\cdot K/2$ assignments. In total, there are $4K+[\log_2(K) -2]\cdot K/2$ accesses.

We propose a new and less complex rescaling procedure. The underlying assumptions are as follows.
A certain cumulative count is the sum of a series of symbol counts and if these counts are halved, then the cumulative count is also halved:
 	\begin{align}\label{eq_rescale_count}
		\!h_{\rm k}[j] &= \!\sum_{i=0}^{j-1} h[j]		\; \rightarrow\; 
		 \sum_{i=0}^{j-1} \frac{1}{2}\tdot h[i] = \frac{1}{2}\tdot \sum_{i=0}^{j-1}h[i]	
		                  =		\frac{1}{2}\tdot h_{\rm k}[j]
		\,.
 	\end{align}
Furthermore, since a cumulative count is the sum of the values from the hierarchical array $v[]$, we can conclude that if the cumulative count needs to be halved, then all related entries of $v[]$ can be halved instead:
 	\begin{align}\label{eq_rescale_v}
		h_{\rm k}[j] &= \sum_{i\in S_j} v[i]		\; \rightarrow\; 
		\frac{1}{2}\tdot h_{\rm k}[j] = \frac{1}{2}\tdot \sum_{i\in S_j} v[i]	= \sum_{i\in S_j} \frac{1}{2}\tdot v[i]	
		\;,
 	\end{align}
where $S_j$ is the set of all entries of $v[]$ that contribute to a certain cumulative count $h_{\rm k}[j]$.
\Algo{alg_rescaleNew} shows the corresponding pseudo-code.
\begin{algorithm}
 \caption{\label{alg_rescaleNew}Proposed rescaling procedure for downscaling the cumulative counts $h_{\rm k}[i]$}
 \hfil
  \begin{minipage}[t]{\mywidth}
		\ifdraft
			\setlength{\baselineskip}{12pt} 
		\fi
		\begin{algorithmic}[1]
		\Procedure{reScaleSumCountsBInew}{$v[~], K$}
			\For { $i \gets 1, 2, \dots, K$}\Comment{for all indices}
				\If {(BitAND($i, 0x01$))}
					\State{} \Comment {odd index: $v[]$ is independent on others}
					\State {$v[i] \gets v[i] - (v[i] >> 1)$}
				\Else
					\State {$testV\!al \gets v[i] - (v[i] >> 1)$}
					\Comment{desired new value}
					\State {$j \gets i - 1$}
					\Comment{index of predecessor}
					\State $compareV\!al \gets 0$
					\Comment{sum of $v[]$ entries of predecessor}
					\State $k \gets i$
					\Comment{$k$ determines num. of iterations}
					\Repeat
						\State $compareV\!al \gets compareV\!al + v[j]$
						\State\Comment {accumulate numbers}
						\State $j \gets $BitAND $(j,j-1)$
						\Comment {next index}
						\State{$k \gets k >> 1$}
					\Until {(BitAND $(k, 0x01) > 0$)}\Comment{until LSB is equal to one}
					\State $v[i] \gets \max(compareV\!al + 1, testV\!al)$ 
					\State\Comment{must be at least one higher than its predecessor}
				\EndIf
			\EndFor
			\State $totalCount = \;${\scshape getSumCount}($v[], K$)
		\State \Return $v[], totalCount$
		\EndProcedure	
		\end{algorithmic}
	\end{minipage}
\end{algorithm}
All elements of $v[i]$ with odd index directly represent the count $h[i-1]$ of the symbol $s_{i-1}$ and these values can be changed independently of other elements. The chosen calculation of $v[i] \gets v[i] - \lfloor v[i] / 2\rfloor$ ensures that $v[i]$ remains positive. 
For the elements at even positions, the desired half is first calculated ($testV\!al$).
It must then be checked, whether this value is greater than the sum $compareV\!al$ of its predecessor.
If not, then the new value $v[i]$ is set to $compareV\!al + 1$.
The efforts for this new rescaling involves $3K$ accesses and assignments, which is distinctly less than to the Fenwick rescaling algorithm.

However, since the operations of (\ref{eq_rescale_count}) and (\ref{eq_rescale_v}) are performed in integer arithmetic, the corresponding quantization effects may result in slightly different values that those produced by {\scshape reScaleSumCountsBI()}. Therefore, the two functions are not compatible with each other.
	%
	
\section{Investigations and Results}
	%
\subsection{Methodology}
	%
All simulations are based on a range-coder implementation in ANSI-C that is combined with the different algorithms. The programs were run on a notebook with i7-1165G7@2.80GHz processor. 
We measured the execution times using the function \verb+__rdtsc()+, which returns the processor time stamp. This stamp records the number of clock cycles since the last reset \cite{Mic24}.
Two time stamps are taken: one right before the data coding starts (including the preparation of all arrays that are required for the modelling) and directly after all symbols have been processed. The difference between both time stamps is taken as the time measure.
It is important to mention that input and output data are stored in memory. This means that file operations for reading or writing are not included in the time measurements. 

The measured time does not necessarily reflect how many CPU instructions have been spent for the coding process. There are two major issues to consider. Firstly, the CPU(s) can share their time with other (background) processes. And secondly, the time needed depends on the CPU clock rate. Modern CPUs have the capability to boost the clock frequency during intensive workloads. On the other hand, during constantly high workloads like in our simulations, the CPU temperature can get too high and so-called thermal throttling is activated, that is, the CPU frequency is reduced until the temperature has dropped below a certain threshold. The CPU frequency can therefore frequently change and the time measurements are not reliable.

To reduce the influence of background processes, all unnecessary programs were closed, the internet connection was switched off, and the measurements were carried out five times each and the smallest value was selected. The problem with the CPU frequency could be narrowed by setting the maximum processor frequency to 75\% in the Windows 11 energy settings. This led to an almost constant CPU frequency of around 2.0 GHz, which could be verified via the task manager.
	%
\subsection{Static processing mode}
	%
The static mode determines the symbol distribution in a preprocessing step by counting all symbols once. This statistic remains fixed for all symbols to be processed.
First, the influence of the alphabet size on the processing speed of different algorithms from Sections \ref{sec_conventionalSearch} to \ref{sec_tableBased} is investigated. Additionally, the procedure {\scshape getSymbolBI()} is tested to show the impact of the binary indexing on the decoding process.

The tests use ten data sequences, each comprising $10^8$ symbols. The sequences differ in the number of the alphabet size, which is chosen as $K=2^n$ with $n=1,2,\dots, 10$.

As the algorithm performance usually depends on the symbol distribution, two typical variants are taken into account: flat and geometric distribution.
	%
\subsubsection{Flat distribution}
\begin{figure*}
	\hfil \includegraphics[scale = 1.2]{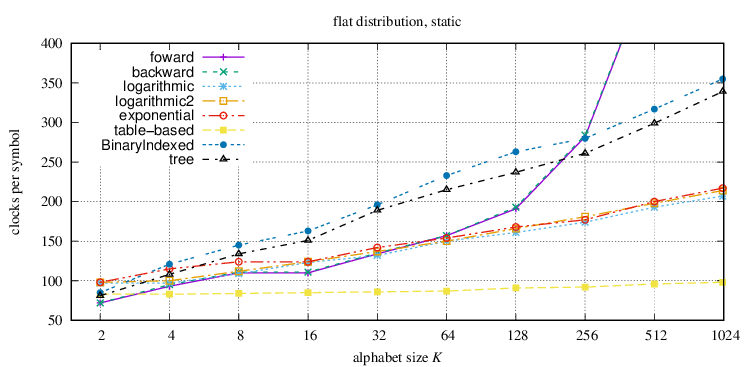}
	\caption{\label{fig_clocksPerSymbolStaticFlat}Required processor clock cycles per symbol at decoder dependent on the alphabet size: flat distribution, $h(s_i) = h(s_j) \forall (i,j)$}
\end{figure*}

	%
In this experiment, the symbol distribution is uniform for all alphabet sizes.
\Figu{fig_clocksPerSymbolStaticFlat} depicts the measured clocks per symbol as function of the alphabet size.
It can clearly be seen that the table-based approach is, as expected, almost independent of the number of different symbols, since only a single access to the table is needed.
Both variants of linear search behave identically, because the average number of steps is always equal to $K/2$. Despite the fact that linear search accesses more elements of $h_{\rm k}[\cdot]$, it is even faster than logarithmic search for small alphabets. Presumably, the higher number of needed additions, comparisons, and variable assignments prevents faster processing in logarithmic search. Only for alphabet sizes larger than about 32 does the higher $\mathcal{O}$-complexity of linear search become a limiting factor.

Since the abscissa axis is on logarithmic scale, the curve for logarithmic search shows a constant ascent of clocks per symbol.
The variants of logarithmic search (logarithmic2 and exponential) cannot positively influence the speed because the distribution is not skewed.
The functions {\scshape getSymbolBI()} (label `BinaryIndexed') and {\scshape getSymbolTree()} (label `tree') also have logarithmic behaviour, but with a higher constant factor. This is probably mainly due to the time-costly access to the memory arrays $v[]$, $le\!f\!t[]$, and $right[]$. The access to the data is less predictable and probably the CPU caching processes become less effective.
	%
\subsubsection{Geometric distribution}
	%
The same experiment was performed for symbols following a truncated geometric distribution:
		\begin{align}
			p(s_i) = \frac{(1-p)\cdot p^i}{1-p^K} \,,\;\; p = 2^{-1/2^k}, \;\;  i=0, 1, 2, \dots, K-1
			\,.
		\end{align}
This type of distribution has already been used by others, e.g. \cite{Sai04}, and is representative for a large class of data.

\begin{figure}
    \captionsetup[subfloat]{labelfont=small,textfont=scriptsize}
	\centering
	\subfloat[]{\includegraphics[scale = 0.9]{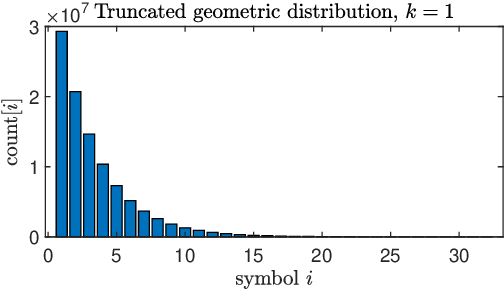}
	\label{fig_geometricDistribution_32}}
\hfil
	\subfloat[]{\includegraphics[scale = 0.9]{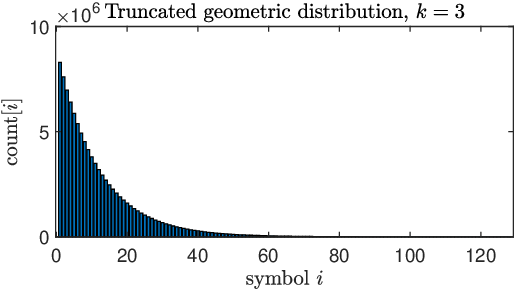}
	\label{fig_geometricDistribution_128}}
	\caption{\label{fig_geometricDistribution}Histograms of $10^8$ symbols $s_i\in\{0,1,2,\dots, K-1\}$ based on a truncated geometric distribution: (a) $K=32$; (b) $K=128$.}
\end{figure}

The parameter $k$ was selected such that the probability of symbols decreases significantly with increasing $i$ and the symbol counts remain above zero:
		\begin{align}
			k(K) = max(0, \lfloor\log_2(K)\rfloor - 4)
			\;.
		\end{align}
\Figu{fig_geometricDistribution} depicts two example histograms for $K=32$ and $K=128$ different symbols.

\begin{figure*}
	\hfil \includegraphics[scale = 1.2]{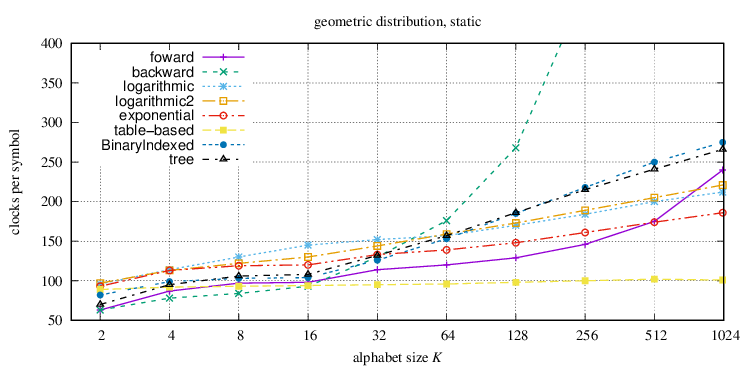}
	\caption{\label{fig_clocksPerSymbolStaticGeometric}Required processor clock cycles per symbol at decoder dependent on the alphabet size $K$ for geometric symbol distributions.}
\end{figure*}

The results of the corresponding time measurement is shown in \Figu{fig_clocksPerSymbolStaticGeometric}.
Now, the forward linear search benefits from symbols with small indices dominating the sequence. Up to $K=256$, it is faster than logarithmic search methods for this particular symbol distribution.   

Backward linear search has to access significantly more than $K/2$ interval boundaries in most cases and accordingly becomes distinctly slower as the alphabet size increases.
Interestingly, it is slightly faster than forward linear search for small $K$. The reason for this is not yet known. It could be that compiler optimizations lead to faster executable code.

We can also see that exponential search can benefit from the higher probability of small symbol indices, leading to faster processing compared to ordinary logarithmic search, while the optimization of the range split (logarithmic2) does not have the desired effect.

The binary indexing and the tree-based scheme also benefit from small alphabets, but remain inferior for $K>100$.

In general, it is recommended to use table-based search in static mode as long as there is enough memory available to store all $h_{\rm k}[K]$ mappings.
	%
\subsection{Adaptive processing mode}
	%
Using the same data sets, the investigation were repeated, running the encoding/decoding software in adaptive mode. That is, the software starts with a flat distribution and after processing each symbol, its count is incremented. 
The decoder time now depends not only on the symbol-identification algorithm, but additionally on the algorithm for updating the cumulative counts, while the encoding time is mainly influenced by the latter.

\begin{figure*}
    \captionsetup[subfloat]{labelfont=small,textfont=scriptsize}
	\centering
	\subfloat[]{\includegraphics[scale = 1.1]{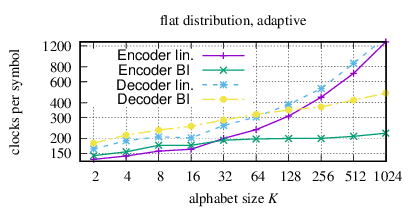}
	\label{fig_clocksPerSymbolAdaptive_flat}}
\hfil
	\subfloat[]{\includegraphics[scale = 1.1]{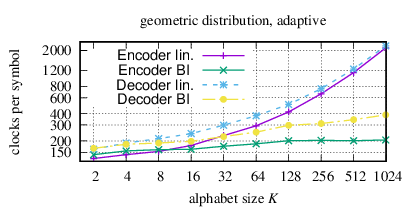}
	\label{fig_clocksPerSymbolAdaptive_geometric}}
	\caption{\label{fig_clocksPerSymbolAdaptive}Adaptive mode: comparison of binary indexing (BI) versus linear updating (encoder and decoder) in combination with logarithmic search (decoder): (a) flat symbol distributions, (b) geometric symbol distributions}
\end{figure*}

\Figu{fig_clocksPerSymbolAdaptive_flat} depicts the results for the encoder and decoder processes when applied to a flat distrbution.
The two solid lines compare the encoder implementations. When using the linear algorithm (`lin.') to update the cumulative counts, the number of clocks per symbol grows rapidly with increasing alphabet size. The binary indexing (`BI') enfolds its superiority at about $K>16$; for larger alphabet sizes, the encoder speed is significantly higher here. 

The decoder implementations generally need more time due to the search-based determination of the symbol to be decoded. The curve `Decoder lin.' is based on linear updating of the cumulative counts ({\scshape updateSumCounts()}) and logarithmic search ({\scshape getSymbolLogarithmic()}), while `Decoder BI' shows the result when using the procedures {\scshape updateSumCountsBI()} and {\scshape getSymbolBI()}. 
Although {\scshape getSymbolBI()} turned out to be distinctly slower than the logarithmic search (see Figure \ref{fig_clocksPerSymbolStaticGeometric}), the faster updating of the cumulative counts more than compensates for this drawback for $K>64$ and the decoding with the binary indexing excels all other variants in terms of speed.

The investigation on geometrically distributed symbols, \Figu{fig_clocksPerSymbolAdaptive_geometric}, shows a clearer advantage for the binary indexing method. Its curves are very similar to those of the flat distribution, but the linear updating of the cumulative counts is now much slower because the symbol indices are smaller on average and more cumulative counts need to be incremented.

In adaptive processing mode, we additionally have to deal with the frequent rescaling of the counts.
Rescaling is not only required when MAX\_TOTALCOUNT is reached. In practical applications, it can be observed that the statistics do change within the data. To adapt to this variation, it is sensible to introduce a kind of `forgetting' by regularly downscaling the counts.

\begin{figure}
	\hfil \includegraphics[scale = 1.2]{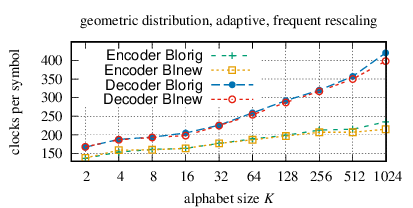}
	\caption{\label{fig_clocksPerSymbolAdaptiveGeometric_r1024}Comparison of rescale algorithms {\scshape reScaleSumCountsBI()} (BIorig) and {\scshape reScaleSumCountsBInew()} (BInew) with a rescaling interval of 1024}
\end{figure}

\Figu{fig_clocksPerSymbolAdaptiveGeometric_r1024} compares the processing times between the rescaling function of Fenwick ({\scshape reScaleSumCountsBI()}) and the proposed procedure {\scshape reScaleSumCountsBInew()}.
Albeit the theoretical complexity metrics indicate a clear advantage for the proposed algorithm, the practical improvement is rather marginal. The main reason may be the fact that rescaling in this study is not applied after processing a single symbol, but after 1024 symbols. It could also be that the proposed algorithm supports the caching of data in the CPU caches less effectively.
	%
\section{Summary}
\label{sec_Summary}\texttt{}
	%
Interval-based coding methods require time-consuming processing steps dominating the total processing times of encoder and decoder.
We investigated several algorithms for their theoretical complexity and carried out practical simulations to verify the performance. The investigations included different alphabet sizes $2\le K\le 1024$ and two modes of operation: static and adaptive.

In static coding mode, only the search for the correct symbol at the decoder is a critical processing step. All variants of logarithmic search outperform the linear search when the alphabet size is sufficiently large. For geometrically distributed symbols, exponential search is the fastest in our comparison. It is only outperformed by the table-based mapping of the code value to a symbol index, which avoids the iterative search process. However, table-based mapping is not applicable in adaptive compression mode.

In the adaptive coding mode, updating the cumulative counts is a second critical processing step that affects both decoder and encoder. The standard linear update was compared with a variant of the binary indexing method first proposed in \cite{Fen94}.
The updating process is significantly accelerated and the time saving can even overcompensate the increased efforts for the symbol search at decoder.

In many real-world cases, the adaptive mode benefits from frequent rescaling of symbol counts to keep track of varying statistics. We proposed a new rescaling procedure based on the binary indexing that has a distinct lower complexity than the original procedure from \cite{Fen94}.

In general, the theoretical advantage of special algorithms only comes into effect for alphabets of a certain size and the threshold depends on the actual symbol distribution. The theoretical advantages of particular algorithms cannot always be transferred into faster execution.
This discrepancy between theoretical and practical performance can mainly be reasoned by the memory management in modern computers. Memory access can become a limiting factor. If source code contains (unpredictable) branches, it is difficult or impossible to prefetch data. This indirectly leads to more cache misses.
Linear algorithms are highly predictable and CPU caches or even CPU registers can be used effectively.
For example, algorithms like {\scshape getSymbolLinForward()} access adjacent memory element in a very predicable manner and take less time to execute for small alphabet sizes than {\scshape getSymbolLogarithmic()}, although the latter requires fewer memory accesses.

\bibliography{literature} 
\bibliographystyle{IEEEtran}

\newpage

\begin{IEEEbiographynophoto}{Tilo Strutz}
holds a Dipl.-Ing. in Electrical Engineering (1994), Dr.-Ing. in Signal Processing (1997) and Dr.-Ing. habil. in Communications Engineering (2002) from the University of Rostock, Germany, where he mainly worked on problems of wavelet-based image compression.
From 2003 to 2007, he worked at the European Molecular Biology Laboratory (Hamburg branch) in the field of multidimensional signal processing and data analysis. He was then Professor of Information and Coding Theory at the Leipzig University of Telecommunications (HfTL) until 2022. Tilo Strutz is now Professor of Image Processing and Computer Vision at Coburg University of Applied Sciences. His research interests range from general signal processing and special problems in image processing to data compression and machine learning.
\end{IEEEbiographynophoto}

\begin{IEEEbiographynophoto}{Nico Schreiber}
studies computer science at Coburg University of Applied Sciences. As a team member and chief programmer of the ‘RoboFreaks’ team, he won the title at the European Championship of the ‘First Lego League’ in 2022.
His research interests currently focus on implementing algorithms for searching in arrays and search trees. 
\end{IEEEbiographynophoto}

\vfill
\includegraphics{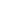}

\end{document}